\documentclass[12pt,english,aps,reprint,showpacs,prb,amsmath,amssymb,floatfix,superscriptaddress,eqsecnum]{revtex4-1}
\usepackage[T1]{fontenc}
\usepackage[latin9]{inputenc}
\usepackage{color}
\usepackage{babel}
\usepackage{bm}
\usepackage{amsmath}
\usepackage{amssymb}
\usepackage{graphicx}
\usepackage[unicode=true,
 bookmarks=true,bookmarksnumbered=false,bookmarksopen=false,
 breaklinks=false,pdfborder={0 0 1},backref=false,colorlinks=false]
 {hyperref}
\hypersetup{pdftitle={Magnetomechanical coupling and ferromagnetic resonance in magnetic nanoparticles},
 pdfauthor={Simon Streib}}

\makeatletter
\bibpunct{[}{]}{,}{n}{}{}
\usepackage{placeins}

\makeatother

\begin{document}

\title{Magnetomechanical coupling and ferromagnetic resonance in magnetic
nanoparticles}

\author{Hedyeh Keshtgar}

\affiliation{Institute for Advanced Studies in Basic Science, 45195 Zanjan, Iran}

\author{Simon Streib}

\affiliation{Kavli Institute of NanoScience, Delft University of Technology, Lorentzweg
1, 2628 CJ Delft, The Netherlands}

\author{Akashdeep Kamra}

\affiliation{Fachbereich Physik, Universität Konstanz, D-78457 Konstanz, Germany}

\author{Yaroslav M. Blanter}

\affiliation{Kavli Institute of NanoScience, Delft University of Technology, Lorentzweg
1, 2628 CJ Delft, The Netherlands}

\author{Gerrit E. W. Bauer}

\affiliation{Kavli Institute of NanoScience, Delft University of Technology, Lorentzweg
1, 2628 CJ Delft, The Netherlands}

\affiliation{Institute for Materials Research and WPI-AIMR, Tohoku University,
Sendai 980-8577, Japan}

\date{March 20, 2017}
\begin{abstract}
We address the theory of the coupled lattice and magnetization dynamics
of freely suspended single-domain nanoparticles. Magnetic anisotropy
generates low-frequency satellite peaks in the microwave absorption
spectrum and a blueshift of the ferromagnetic resonance (FMR) frequency.
The low-frequency resonances are very sharp with maxima exceeding
that of the FMR, because their magnetic and mechanical precessions
are locked, thereby suppressing the effective Gilbert damping. Magnetic
nanoparticles can operate as nearly ideal motors that convert electromagnetic
into mechanical energy. The Barnett damping term is essential for
obtaining physically meaningful results.
\end{abstract}

\maketitle

\section{Introduction}

Magnetic nanoparticles (nanomagnets) are of fundamental interest in
physics by forming a link between the atomic and macroscopic world.
Their practical importance stems from the tunability of their magnetic
properties \cite{Kolhatkar13}, which is employed in patterned media
for high density magnetic data storage applications \cite{Frey11}
as well as in biomedicine and biotechnology \cite{Pankhurst03,Tartaj03,Berry03,Reenen14}.
Superparamagnetic particles are used for diagnostics, stirring of
liquids, and magnetic tweezers \cite{Oene15}. The heat generated
by the magnetization dynamics under resonance conditions is employed
for hyperthermia cancer treatment \cite{Nandori12,Vallejo13,Janosfalvi14}.
Molecular based magnets can cross the border from the classical into
the quantum regime \cite{Kahn00,Chudnovsky14}. The magnetic properties
of individual atomic clusters can be studied by molecular beam techniques
\cite{Bucher91,Billas93,Ma14}.

Einstein, de Haas, and Barnett \cite{Einstein15,Barnett15} established
the equivalence of magnetic and mechanical angular momentum of electrons
by demonstrating the coupling between magnetization and global rotations.
Spin and lattice are also coupled by magnetic anisotropy, induced
either by dipolar forces or crystalline fields. A quite different
interaction channel is the magnetoelastic coupling between lattice
waves (phonons) and spin waves (magnons) with finite wave vectors.
This magnetoelastic coupling between the magnetic order and the underlying
crystalline lattice has been explored half a century ago by Kittel
\cite{Kittel58} and Comstock \cite{Comstock62a,Comstock62b}. The
coupling between spin and lattice causes spin relaxation including
Gilbert damping of the magnetization dynamics \cite{Gilbert04,Mohanty04}. 

``Spin mechanics'' of thin films and nanostructures encompasses
many phenomena such as the actuation of the magnetization dynamics
by ultrasound \cite{Uchida11,Weiler12,Labanowski16}, the dynamics
of ferromagnetic cantilevers \cite{Kovalev03,Kovalev05,Kovalev06},
spin current-induced mechanical torques \cite{Mohanty04,Malshukov05},
and rotating magnetic nanostructures \cite{Bretzel09}. The Barnett
effect by rotation has been observed experimentally by nuclear magnetic
resonance \cite{Chudo14}. The coupled dynamics of small magnetic
spheres has been studied theoretically by Usov and Liubimov \cite{Usov15}
and Rusconi and Romero-Isart \cite{Rusconi16} in classical and quantum
mechanical regimes, respectively. A precessing single-domain ferromagnetic
needle is a sensitive magnetometer \cite{Kimball16}, while a diamagnetically
levitated nanomagnet can serve as a sensitive force and inertial sensor
\cite{Prat-Camps17}. A stabilization of the quantum spin of molecular
magnets by coupling to a cantilever has been predicted \cite{Kovalev11,Garanin11}
and observed recently \cite{Ganzhorn16}.

Here we formulate the dynamics of rigid and single-domain magnetic
nanoparticles with emphasis on the effects of magnetic anisotropy
and shape. We derive the equations of motion of the macrospin and
macrolattice vectors that are coupled by magnetic anisotropy and Gilbert
damping. We obtain the normal modes and microwave absorption spectra
in terms of the linear response to ac magnetic fields. We demonstrate
remarkable changes in the normal modes of motion that can be excited
by microwaves. We predict microwave-activated nearly undamped mechanical
precession. Anisotropic magnetic nanoparticles are therefore suitable
for studies of non-linearities, chaos, and macroscopic quantum effects. 

In Sec.~\ref{sec:Macrospin-model} we introduce the model of the
nanomagnet and give an expression for its energy. In Sec.~\ref{sec:LLG derivation}
we discuss Hamilton's equation of motion for the magnetization of
a freely rotating particle, which is identical to the Landau-Lifshitz
equation. We then derive the coupled equations of motion of magnetization
and lattice in Sec.~\ref{sec:Equations-of-motion}. Our results for
the easy-axis and easy-plane configurations are presented in Secs.~\ref{sec:Easy-axis-configuration}
and \ref{sec:Easy-plane-Configuration}. We discuss and summarize
our results in Secs.~\ref{sec:Discussion} and \ref{sec:Summary}.
In the Appendices \ref{sec:Coordinate-systems} to \ref{sec:FMR-absorption}
we present additional technical details and derivations.

\section{Macrospin model\label{sec:Macrospin-model}}

\begin{figure}
\begin{centering}
\includegraphics{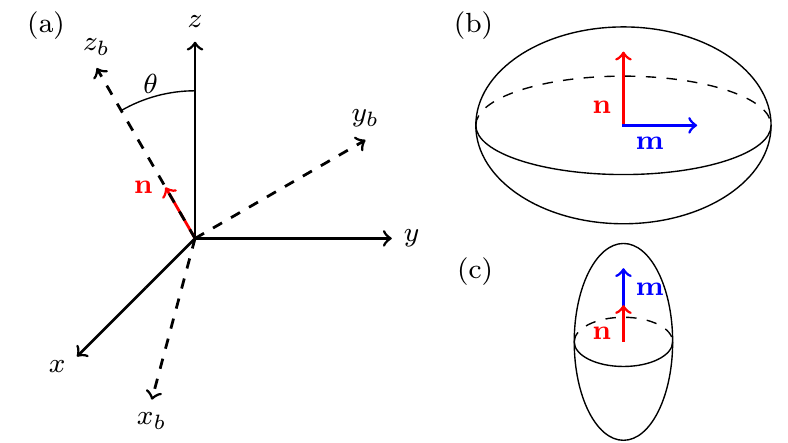}
\par\end{centering}
\caption{(a) Laboratory frame ($x$, $y$, $z$) and (moving) body frame ($x_{b}$,
$y_{b}$, $z_{b}$) of a nanomagnet with principal axis $\mathbf{n}$
along the $z_{b}$-axis. The directions of $\mathbf{n}$ and magnetization
$\mathbf{m}$ are shown for (b) oblate and (c) prolate spheroids with
dipolar magnetic anisotropy.\label{fig:coordinate frames}}
\end{figure}
We consider a small isolated nanomagnet that justifies the macrospin
and macrolattice approximations, in which all internal motion is adiabatically
decoupled from the macroscopic degrees of freedom, rendering the magnetoelastic
coupling irrelevant. 

We focus on non-spherical nanoparticles with mass density $\rho\left(\boldsymbol{r}\right)$
and tensor of inertia
\begin{equation}
\mathcal{I}=\int d^{3}r\,\rho({\bf r})\left[\left({\bf r}\cdot{\bf r}\right)\hat{1}-{\bf r}\otimes{\bf r}\right],\label{eq:ToI}
\end{equation}
where $\hat{1}$ is the 3x3 unit matrix. The mechanical properties
of an arbitrarily shaped rigid particle is identical to that of an
ellipsoid with a surface that in a coordinate system defined along
the symmetry axes (in which $\mathcal{I}$ is diagonal) reads
\begin{equation}
\left(\frac{x}{c}\right)^{2}+\left(\frac{y}{b}\right)^{2}+\left(\frac{z}{a}\right)^{2}=1,
\end{equation}
where $a,b,c$ are the shape parameters (principal radii). The volume
is $V=4\pi abc/3$, total mass $Q=\rho V$, and principal moments
of inertia $I_{1}=Q\left(a^{2}+b^{2}\right)/5,$ $I_{2}=Q\left(a^{2}+c^{2}\right)/5,$
$I_{3}=Q\left(b^{2}+c^{2}\right)/5$. We focus in the following on
prolate $(a>b=c)$ and oblate $(a<b=c)$ spheroids, because this allows
analytic solutions of the dynamics close to the minimum energy state. 

We assume that the particle is smaller than the critical size $d_{\mathrm{cr}}\sim36\sqrt{AK_{A}}/(\mu_{0}M_{s}^{2})$
for magnetic domain formation \cite{Coey09}, where $A$ is the exchange
constant, $K_{A}$ the anisotropy constant, $M_{s}$ the saturation
magnetization, and $\mu_{0}=4\pi\times10^{-7}\mathrm{N}\,\mathrm{A}^{-2}$
the vacuum permeability. For strong ferromagnets these parameters
are typically in the range $A\in\left[5,30\right]\;\mathrm{pJ\,m^{-1}}$,
$K_{A}\in\left[10,20000\right]\;\mathrm{kJ\,m^{-3}}$, $M_{s}\in\left[0.4,1.7\right]\;\mathrm{MA\,m^{-1}}$,
leading to $d_{\mathrm{cr}}\in\left[1,500\right]\;\mathrm{nm}$ \cite{Coey09}.
For a spherical particle of radius $R$ with sound velocity $v$,
the lowest phonon mode frequency is approximately \cite{Romero11}
\begin{equation}
\frac{\omega_{\mathrm{ph}}}{2\pi}\approx\frac{v}{4R}=0.25\left(\frac{v/(10^{3}\mathrm{\frac{m}{s}})}{R/\mathrm{nm}}\right)\mathrm{THz},
\end{equation}
while the lowest magnon mode (for bulk dispersion relation $\hbar\omega_{mag}=Dk^{2}$)
\begin{equation}
\frac{\omega_{\mathrm{mag}}}{2\pi}\approx\frac{\pi D}{8\hbar R^{2}}=0.6\left(\frac{D/(\mathrm{meV\,nm^{2})}}{R^{2}/\mathrm{nm}^{2}}\right)\mathrm{THz},
\end{equation}
where the spin wave stiffness $D=2g\mu_{B}A/M_{s}$ is typically of
the order $\mathrm{meV\,nm^{2}}$ \cite{Coey09}, e.g., $D=2.81\;\mathrm{meV\,nm^{2}}$
for iron \cite{Shirane68}. We may disregard spin and lattice waves
and the effects of their thermal fluctuations when the first excited
modes are at sufficiently higher frequencies than that of the total
motion (the latter is typically in the GHz range) and therefore adiabatically
decoupled \cite{Romero11,Rusconi16}, i.e. the macrospin and macrolattice
model is valid. Thermal fluctuations of the magnetization with respect
to the lattice do not play an important role below the blocking temperature,
$T_{B}\sim K_{A}V/(25k_{B})$ \cite{Knobel08}, where $k_{B}$ is
the Boltzmann constant. For $k_{B}T\ll VM_{s}\mu_{0}H_{0}$, thermal
fluctuations of the magnetization with respect to the static external
magnetic field $H_{0}$ are suppressed.

Under the conditions stipulated above the classical dynamics (disregarding
translations of the center of mass) is described in terms of the magnetization
vector $\mathbf{M}=M_{s}\mathbf{m}$ (with $|\mathbf{m}|=1$) and
the three Euler angles ($\theta,\phi,\psi$) of the crystal orientation
direction in terms of the axis $\mathbf{n}(\theta,\phi)$ and a rotation
angle $\psi$ around it (see Appendix \ref{sec:Coordinate-systems}
for details). The total energy can be split up into several contributions,

\begin{equation}
E=E_{T}+E_{Z}+E_{D}+E_{K}.\label{eq:energy}
\end{equation}
$E_{T}=\frac{1}{2}\mathbf{\boldsymbol{\Omega}}^{\mathrm{T}}\mathcal{I}\mathbf{\boldsymbol{\Omega}}$
is the kinetic energy of the rotational motion of the nanomagnet in
terms of the angular frequency vector $\mathbf{\boldsymbol{\Omega}}$.
$E_{Z}=-\mu_{0}V\mathbf{M}\cdot\mathbf{H}_{\mathrm{ext}}$ is the
Zeeman energy in a magnetic field $\mathbf{H}_{\mathrm{ext}}$. $E_{D}=\frac{1}{2}\mu_{0}V\mathbf{M^{\mathit{\mathrm{T}}}}\mathcal{D}\mathbf{M}$
is the magnetostatic self-energy with particle shape-dependent demagnetization
tensor $\mathcal{D}$. $E_{K}=K_{1}V(\mathbf{m}\times\mathbf{n})^{2}$
is the (uniaxial) magnetocrystalline anisotropy energy, assuming that
the easy axis is along $\mathbf{n}$, and $K_{1}$ is the material-dependent
anisotropy constant. 

We consider an inertial lab frame with origin at the center of mass
and a moving frame with axes fixed in the body. The lab frame is spanned
by basis vectors $\mathbf{e}_{x}$, $\mathbf{e}_{y}$, $\mathbf{e}_{z}$,
and the body frame by basis vectors $\mathbf{e}_{x_{b}}$, $\mathbf{e}_{y_{b}}$,
$\mathbf{e}_{z_{b}}$ (see Fig.~\ref{fig:coordinate frames}). The
body axes are taken to be the principal axes that diagonalize the
tensor of inertia. For spheroids with $b=c$ the inertia and demagnetizing
tensors in the body frame have the form
\begin{equation}
\mathcal{I}_{b}=\begin{pmatrix}I_{\perp} & 0 & 0\\
0 & I_{\perp} & 0\\
0 & 0 & I_{3}
\end{pmatrix},\;\;\;\mathcal{D}_{b}=\begin{pmatrix}D_{\perp} & 0 & 0\\
0 & D_{\perp} & 0\\
0 & 0 & D_{3}
\end{pmatrix},
\end{equation}
with $I_{\perp}=Q\left(a^{2}+b^{2}\right)/5$ and $I_{3}=2Qb^{2}/5$;
the elements $D_{\perp}$and $D_{3}$ for magnetic spheroids are given
in \cite{Osborne45}. The particle shape enters the equations of motion
via $I_{\perp}$, $I_{3}$, and the difference $D_{3}-D_{\perp}$,
the latter reduces to $-1/2$ for a thin needle and $1$ for a thin
disk. When
\begin{equation}
E_{\perp}-E_{\parallel}=K_{A}V=K_{1}V-\frac{1}{2}\mu_{0}VM_{s}^{2}\left(D_{3}-D_{\perp}\right)\label{eq:energy difference}
\end{equation}
is larger than zero, the configuration $\mathbf{m}\parallel\mathbf{n}$
is stable (``easy axis''); otherwise $\mathbf{m}\perp\mathbf{n}$
(``easy plane''). The anisotropy constant $K_{A}$ includes both
magnetocrystalline and shape anisotropy. 

\section{Landau-Lifshitz equation\label{sec:LLG derivation}}

For reference we rederive here the classical equation of motion of
the magnetization. The magnetization of the particle at rest is related
to the angular momentum $\mathbf{S}=-VM_{s}\mathbf{m}/\gamma$, where
$\gamma=1.76\times10^{11}\;\mathrm{s}^{-1}\mathrm{T}^{-1}$ is (minus)
the gyromagnetic ratio of the electron. The Poisson bracket relations
for angular momentum are
\begin{equation}
\left\{ S_{\alpha},S_{\beta}\right\} =\epsilon_{\alpha\beta\gamma}S_{\gamma}.
\end{equation}
Hamilton's equation of motion reads
\begin{equation}
\frac{d}{dt}\mathbf{S}=\left\{ \mathbf{S},\mathcal{H}\right\} ,\label{eq:Hamilton}
\end{equation}
where $\mathcal{H}\equiv E$ is the Hamiltonian. We consider a general
model Hamiltonian of a single macrospin coupled to the macrolattice,
\begin{equation}
\mathcal{H}=\sum_{i,j,k\in\mathbb{N}_{0}}a_{ijk}(\mathbf{n},\mathbf{L})S_{x}^{i}S_{y}^{j}S_{z}^{k},
\end{equation}
where the coefficients $a_{ijk}(\mathbf{n},\mathbf{L})$ may depend
on the orientation $\mathbf{n}$ of the lattice and its mechanical
angular momentum $\mathbf{L}=\mathcal{I}\mathbf{\Omega}$. Since lattice
and magnetization are different degrees of freedom, the Poisson brackets
$\left\{ \mathbf{n},\mathbf{S}\right\} =\left\{ \mathbf{L},\mathbf{S}\right\} =0$
and therefore $\left\{ a_{ijk}(\mathbf{n},\mathbf{L}),\mathbf{S}\right\} =0$.
We derive in Appendix~\ref{sec:Hamilton's-equation}
\begin{equation}
\left\{ \mathbf{S},\mathcal{H}\right\} =\sum_{i,j,k\in\mathbb{N}_{0}}a_{ijk}(\mathbf{n},\mathbf{L})\begin{pmatrix}iS_{x}^{i-1}S_{y}^{j}S_{z}^{k}\\
jS_{x}^{i}S_{y}^{j-1}S_{z}^{k}\\
kS_{x}^{i}S_{y}^{j}S_{z}^{k-1}
\end{pmatrix}\times\mathbf{S},\label{eq:Hamilton evaluated}
\end{equation}
which is the Landau-Lifshitz equation \cite{Landau35},
\begin{equation}
\frac{d}{dt}\mathbf{S}=\left.\mathbf{\nabla}_{\mathbf{S}}\mathcal{H}\right|_{\mathbf{n},\mathbf{L}=\mathrm{const.}}\times\mathbf{S}.\label{eq:LL}
\end{equation}
In accordance with Eq.~(\ref{eq:Hamilton evaluated}), the gradient
in Eq.~(\ref{eq:LL}) has to be evaluated for constant $\mathbf{n}$
and $\mathbf{L}$. 

The rotational kinetic energy $E_{T}=\frac{1}{2}\mathbf{\boldsymbol{\Omega}}^{\mathrm{T}}\mathcal{I}\mathbf{\boldsymbol{\Omega}}$
does not contribute to this equation of motion directly since $\left\{ \mathbf{S},E_{T}\right\} =0$.
However, $E_{T}$ is crucial when considering the energy of the nanomagnet
under the constraint of conserved total angular momentum $\mathbf{J}=\mathbf{L}+\mathbf{S}$.
Minimizing the energy of the nanomagnet under the constraint of constant
$\mathbf{J}$ is equivalent to
\begin{equation}
\tilde{\mathbf{H}}_{\mathrm{eff}}=\left.-\frac{1}{\mu_{0}VM_{s}}\mathbf{\nabla}_{\mathbf{m}}E\right|_{\mathbf{J}=\mathrm{const.}}=0,
\end{equation}
where the rotational kinetic energy $E_{T}$ contributes the Barnett
field
\begin{equation}
\mathbf{H}_{B}=\left.-\frac{1}{\mu_{0}VM_{s}}\mathbf{\nabla}_{\mathbf{m}}E_{T}\right|_{\mathbf{J}=\mathrm{const.}}=-\frac{\mathbf{\boldsymbol{\Omega}}}{\gamma\mu_{0}},
\end{equation}
which gives rise to the Barnett effect (magnetization by rotation)
\cite{Barnett15}. Although the Barnett field appears here in the
effective field $\tilde{\mathbf{H}}_{\mathrm{eff}}$ when minimizing
the energy, it is not part of the effective field $\mathbf{H}_{\mathrm{eff}}$
of the Landau-Lifshitz equation, 
\begin{equation}
\mathbf{H}_{\mathrm{eff}}=\left.-\frac{1}{\mu_{0}VM_{s}}\mathbf{\nabla}_{\mathbf{m}}E\right|_{\mathbf{n},\mathbf{L}=\mathrm{const.}},\label{eq:H_eff LL}
\end{equation}
where $\mathbf{L}$ is kept constant instead of $\mathbf{J}$. In
the Landau-Lifshitz-Gilbert equation in the laboratory frame the Barnett
effect operates by modifying the Gilbert damping torque as shown below.

\section{Equations of motion\label{sec:Equations-of-motion}}

We now derive the coupled equations of motion of the magnetization
$\mathbf{m}$ and the Euler angles ($\phi,\theta,\psi$). The magnetization
dynamics is described by the Landau-Lifshitz-Gilbert equation \cite{Landau35,Gilbert04}
\begin{eqnarray}
\dot{\mathbf{m}} & = & -\gamma\mu_{0}\mathbf{m}\times\mathbf{H}_{\mathrm{eff}}+\boldsymbol{\mathbf{\tau}}_{m}^{(\alpha)},\label{eq:LLG equation}
\end{eqnarray}
where the effective magnetic field Eq.~(\ref{eq:H_eff LL}) follows
from the energy Eq.~(\ref{eq:energy}),
\begin{equation}
\mathbf{H}_{\mathrm{eff}}=\mathbf{H}_{\mathrm{ext}}+\mathbf{H}_{D}+\mathbf{H}_{K},\label{eq:Heff}
\end{equation}
and $\mathbf{\tau}_{m}^{(\alpha)}$ is the (Gilbert) damping torque.
The external magnetic field ${\bf H}_{\mathrm{ext}}$ is the only
source of angular momentum; all other torques acting on the total
angular momentum ${\bf J}={\bf L}-VM_{s}{\bf m/\gamma}$ cancel. From
\begin{equation}
\dot{{\bf J}}=\mu_{0}VM_{s}\mathbf{m}\times{\bf H}_{\mathrm{ext}},
\end{equation}
we obtain the mechanical torque as time-derivative of the mechanical
angular momentum, which leads to Newton's Law
\begin{equation}
\dot{\mathbf{L}}=\frac{VM_{s}}{\gamma}\dot{\mathbf{m}}+\mu_{0}VM_{s}\mathbf{m}\times{\bf H}_{\mathrm{ext}}.\label{eq:mechanical torque}
\end{equation}
The dissipation parameterized by the Gilbert constant \cite{Gilbert04}
damps the relative motion of magnetization and lattice. In the body
frame of the lattice \cite{Bretzel09}
\begin{equation}
\mathbf{\boldsymbol{\tau}}_{m,b}^{(\alpha)}=\alpha\mathbf{m}_{b}\times\dot{\mathbf{m}}_{b},
\end{equation}
where the subscript $b$ indicates vectors in the body frame. Transformed
into the lab frame (see Appendix~\ref{sec:Coordinate-systems})
\begin{equation}
\boldsymbol{\tau}_{m}^{(\alpha)}=\alpha\left[\mathbf{m}\times\dot{\mathbf{m}}+\mathbf{m}\times\left(\mathbf{m}\times\mathbf{\boldsymbol{\Omega}}\right)\right].
\end{equation}
This torque is an angular momentum current that flows from the magnet
into lattice \cite{Mohanty04}. Angular momentum is conserved, but
the generated heat is assumed to ultimately be radiated away. In vacuum
there is no direct dissipation of the rigid mechanical dynamics.

The Barnett field $\mu_{0}\mathbf{H}_{B}=-\mathbf{\boldsymbol{\Omega}}/\gamma$
enters in the lab frame only in the damping term $\boldsymbol{\tau}_{m}^{(\alpha)}$.
To leading order in $\alpha$
\begin{equation}
\dot{\mathbf{m}}\approx-\gamma\mu_{0}\mathbf{m}\times\mathbf{H}_{\mathrm{eff}}-\alpha\gamma\mu_{0}\mathbf{m}\times\left[\mathbf{m}\times\left(\mathbf{H}_{\mathrm{eff}}+\mathbf{H}_{B}\right)\right]+\mathcal{O}(\alpha^{2}).
\end{equation}
The contribution of $\mathbf{H}_{B}$ in the damping term causes the
Barnett effect \cite{Barnett15}. We find that this Barnett damping
is very significant for the coupled dynamics even though no fast lattice
rotation is enforced: without Barnett damping the FMR absorption of
the low-frequency modes described below would become negative. 

\section{Easy-axis configuration\label{sec:Easy-axis-configuration}}

\begin{figure}
\begin{centering}
\includegraphics{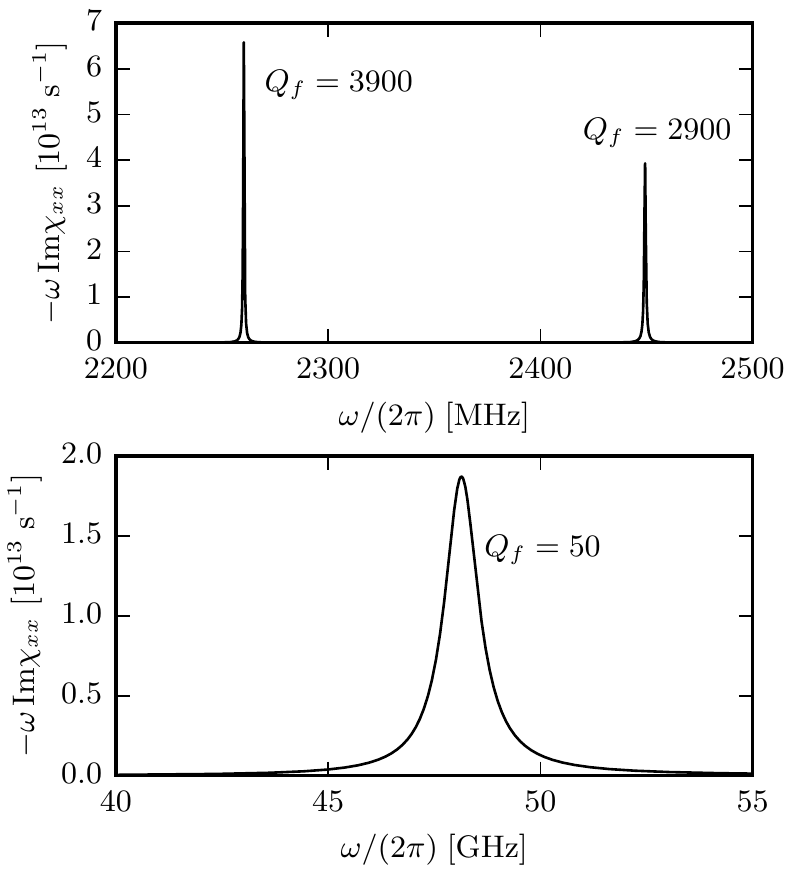}
\par\end{centering}
\caption{Low- and high-frequency resonances in the FMR spectrum of an Fe nanosphere
of $2\;\mathrm{nm}$ diameter in a static magnetic field of $0.65\;\mathrm{T}$
with Gilbert damping constant $\alpha=0.01$; quality factor $Q_{f}=\omega/\mathrm{(2\eta)}$.
\label{fig:iron sphere 0.1T}}
\end{figure}

We first consider an easy-axis configuration ($\mathbf{m}\parallel\mathbf{n}\parallel\mathbf{\mathbf{\mathbf{e}_{\mathit{z}}}}$)
in the presence of an external magnetic field with a large dc component
$H_{0}$ along $\mathbf{\mathbf{\mathbf{e}_{\mathit{z}}}}$ and a
small transverse ac component, $\mathbf{H}_{\mathrm{ext}}=\begin{pmatrix}h_{x}(t), & h_{y}(t), & H_{0}\end{pmatrix}^{\mathrm{T}}$,
with $h_{x}(t)\propto h_{y}(t)\propto e^{i\omega t}.$ Linearizing
the equations of motion in terms of small transverse amplitudes, we
can solve (\ref{eq:LLG equation}) and (\ref{eq:mechanical torque})
analytically to obtain the linear response to $\mathbf{h}$ (see Appendix
\ref{sec:Linearized-equations} for the derivation), i.e. the transverse
magnetic susceptibility. Since we find $\dot{\Omega}_{z}=0$, we disregard
an initial net rotation by setting $\Omega_{z}=0$. For small damping
$\alpha\ll1$, the normal modes are given by the positive solutions
of the equations 
\begin{equation}
\omega^{3}\mp\omega^{2}\omega_{0}-\omega\omega_{c}\omega_{A}\pm\omega_{c}\omega_{A}\omega_{H}=0,\label{eq:normal mode equation}
\end{equation}
where $\omega_{H}=\gamma\mu_{0}H_{0}$, $\omega_{A}=2\gamma K_{A}/M_{s}$,
$\omega_{0}=\omega_{H}+\omega_{A}$, and $\omega_{c}=M_{s}V/(\gamma I_{\perp})$
is the natural mechanical frequency governed by the spin angular momentum.
Note that the equivalent negative solutions of Eq.~(\ref{eq:normal mode equation})
have the same absolute values as the positive solutions. We find that
the FMR mode $\omega_{0}$ is blueshifted to $\omega_{\parallel}=\omega_{0}+\delta\omega_{\parallel}$
with 
\begin{equation}
\delta\omega_{\parallel}\approx\frac{\omega_{A}^{2}\omega_{c}}{\omega_{0}^{2}}>0,
\end{equation}
which is significant for small nanomagnets with large saturation magnetization
and low mass density. It is a counterclockwise precession of $\mathbf{m}$
with $\mathbf{n}$ nearly at rest. 

\begin{figure}
\begin{centering}
\includegraphics{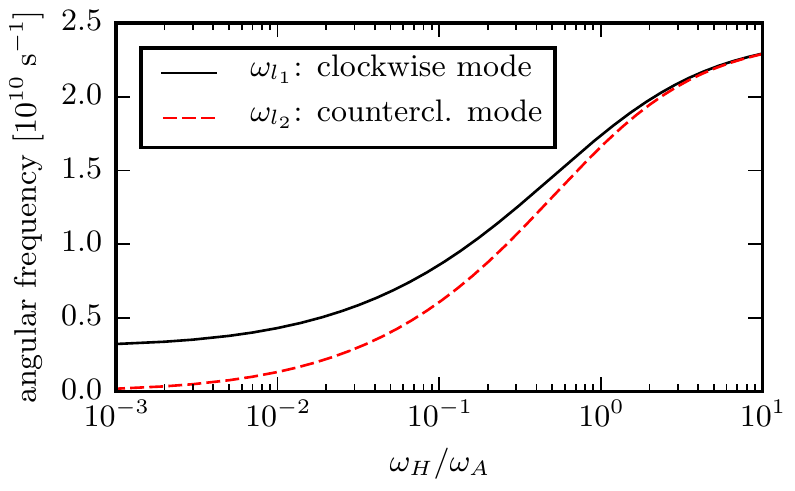}
\par\end{centering}
\caption{Low-frequency magnetomechanical modes $\omega_{l_{1}}$ and $\omega_{l_{2}}$
of an Fe nanosphere of $2\;\mathrm{nm}$ diameter.\label{fig:modes}}
\end{figure}

Two additional low-frequency modes emerge. For $\omega\ll\omega_{0},\omega_{A}$
we may disregard the cubic terms in Eq.~(\ref{eq:normal mode equation})
and find 

\begin{equation}
\omega_{l_{1,2}}\approx\sqrt{\left(\frac{\omega_{c}\omega_{A}}{2\omega_{0}}\right)^{2}+\frac{\omega_{H}\omega_{c}\omega_{A}}{\omega_{0}}}\pm\frac{\omega_{c}\omega_{A}}{2\omega_{0}}.\label{eq:low freq modes}
\end{equation}
At low frequencies, the magnetization can follow the lattice nearly
adiabatically, so these modes correspond to clockwise and counterclockwise
precessions of nearly parallel vectors $\mathbf{m}$ and \textbf{$\mathbf{n}$},
but with a phase lag that generates the splitting.\emph{ }The frequency
of the clockwise mode $\omega_{l_{1}}>\omega_{l_{2}}$ (see Fig.~\ref{fig:modes}).
Since magnetization and mass precess in unison, the effective Gilbert
damping is expected to be strongly suppressed as observable in FMR
absorption spectra as shown below. 

The absorbed FMR power is (see Appendix \ref{sec:FMR-absorption})
\begin{equation}
P=-\frac{\mu_{0}V}{2}\omega\mathrm{Im}\left(\mathbf{h}_{\perp}^{*\mathrm{T}}\chi{\bf h}_{\perp}\right),
\end{equation}
where $\mathbf{h}_{\perp}$ is the ac field normal to the static magnetic
field $H_{0}\mathbf{\mathbf{e}_{\mathit{z}}}$ and 
\begin{equation}
\chi_{\alpha\beta}=\left.\frac{M_{\alpha}}{h_{\beta}}\right|_{\mathbf{h}_{\perp}=0}
\end{equation}
is the transverse magnetic susceptibility tensor ($\alpha,\beta=x,y$).
The diagonal ($\chi_{xx}=\chi_{yy}$) and the off-diagonal components
($\chi_{xy}=-\chi_{yx}$) both contribute to the absorption spectrum
near the resonance frequencies, $\left|\mathrm{Im}\chi_{xx}\right|\approx\left|\mathrm{Re}\chi_{xy}\right|$.\emph{
}For $\alpha\ll1$, we find that the sum rule 
\begin{equation}
\int_{0}^{\infty}d\omega\,\left(-\omega\mathrm{Im}\chi_{xx}(\omega)\right)\approx\frac{\pi}{2}\omega_{0}\omega_{M},
\end{equation}
where $\omega_{M}=\gamma\mu_{0}M_{s}$, does not depend on $\omega_{c}$,
meaning that the coupling does not generate oscillator strengths,
only redistributes it. Close to a resonance
\begin{equation}
-\omega\mathrm{Im}\chi_{xx}(\omega)\sim F\frac{\eta^{2}}{(\omega-\omega_{i})^{2}+\eta^{2}},
\end{equation}
with integral $\pi\eta F$. For the low-frequency modes the maximum
$F\sim\frac{1}{2}\omega_{M}\omega_{A}^{2}/(\alpha\omega_{H}^{2})$
with broadening $\eta\sim\frac{1}{2}\alpha\omega_{c}\omega_{H}^{2}/(\omega_{A}+\omega_{H})^{2}$;
for the FMR mode $F\sim\frac{1}{2}\omega_{M}/\alpha$ with $\eta\sim\alpha\omega_{0}$.

Let us consider an iron sphere with $2\;\mathrm{nm}$ diameter ($a=b=1\;\mathrm{nm}$)
under $\mu_{0}H_{0}=0.65\;\mathrm{T}$ or $\omega_{H}/(2\pi)=18.2\;\mathrm{GHz}$.
Its magnetization $\omega_{M}/(2\pi)=60.33\;\mathrm{GHz}$, crystalline
anisotropy $\omega_{A}/(2\pi)=29.74\;\mathrm{GHz}$ \cite{Park00},
and the magnetomechanical coupling $\omega_{c}/(2\pi)=0.5(\mathrm{nm}/a)^{2}\;\mathrm{GHz}$.
The blocking temperature is $T_{B}\sim11(a/\mathrm{nm})^{3}\;\mathrm{K}$
and $|E_{Z}|/(k_{B}T_{B})\approx30$, while the critical size for
domain formation $d_{\mathrm{cr}}\sim20\;\mathrm{nm}$ \cite{Butler75,Muxworthy15}.
We adopt a typical Gilbert damping constant $\alpha=0.01$. The calculated
FMR spectra close to the three resonances are shown in Fig.~\ref{fig:iron sphere 0.1T}.
Both low-frequency resonances are very sharp with a peak value up
to 3.5 times larger than that of the high-frequency resonance, although
the integrated intensity ratio is only 0.2 \%. Long relaxation times
of low-frequency modes that imply narrow resonances have been predicted
for spherical nanomagnets \cite{Usov15}. The blueshift of the high-frequency
resonance is $\delta\omega_{\parallel}/(2\pi)\approx0.2(\mathrm{nm}/a)^{2}\;\mathrm{GHz}$.
In Fig.~\ref{fig:modes} we plot the low-frequency modes $\omega_{l_{1}}$
and $\omega_{l_{2}}$ as a function of $\omega_{H}/\omega_{A}$. For
$\omega_{H}/\omega_{A}\to0$, $\omega_{l_{1}}\approx\omega_{c}$ and
$\omega_{l_{2}}\to0$. The low-frequency modes become degenerate in
the limit $\omega_{H}/\omega_{A}\to\infty$. 

In $\varepsilon$-Fe$_{2}$O$_{3}$ \cite{Ohkoshi15} magnetization
is reduced, resulting in $\omega_{M}/(2\pi)=2.73\;\mathrm{GHz}$ and
$\omega_{c}/(2\pi)=35(\mathrm{nm/\mathit{a})^{2}}\;\mathrm{MHz}$.
 For the single-molecule magnet $\mathrm{TbPc_{2}}$ \cite{Ganzhorn16},
we estimate $\omega_{A}/(2\pi)\sim5\;\mathrm{THz}$ \cite{Ishikawa03},
$\omega_{M}/(2\pi)\sim10\;\mathrm{GHz}$, $\omega_{c}/(2\pi)\sim100\;\mathrm{MHz}$
\cite{Ganzhorn13}, giving access to the strong-anisotropy regime
with ultra-low effective damping.

\section{Easy-plane Configuration\label{sec:Easy-plane-Configuration}}

An easy-plane anisotropy aligns the equilibrium magnetization normal
to the principal axis ($\mathbf{m}\perp\mathbf{n}$), which is typically
caused by the shape anisotropy of pancake-like oblate spheroids corresponding
to $\omega_{A}<0$. We choose an external magnetic field with a static
component in the plane $H_{0}\mathbf{\mathbf{e}}_{y}$ and an ac field
along $x$ and $z$, while the equilibrium $\mathbf{n}$ points along
$\mathbf{e}_{z}$ (see Fig. 1(b)). For $\theta\ll1$, $m_{y}\approx1$,
$n_{z}\approx1$, we again obtain analytic solutions for $\mathbf{m}$
and $\mathbf{n}$ (see Appendix \ref{sec:Linearized-equations}).
We find two singularities in the magnetic susceptibility tensor with
frequencies (for $\alpha\ll1$)
\begin{eqnarray}
\omega_{\perp} & \approx & \omega_{H}\sqrt{1-\frac{\omega_{A}}{\omega_{H}}-\frac{\omega_{c}\omega_{A}}{\omega_{H}^{2}}},\\
\omega_{l} & \approx & \sqrt{\frac{\omega_{H}^{2}\omega_{c}\omega_{A}}{\omega_{A}\omega_{H}-\omega_{H}^{2}+\omega_{c}\omega_{A}}}.
\end{eqnarray}
Since $n_{x}$ does not depend on time there is only one low-frequency
mode $\omega_{l}$, viz. an oscillation about the $x$-axis of the
nanomagnet. Linearization results in $\dot{L}_{y}\approx VM_{s}\dot{m}_{y}/\gamma\approx0$
and implies $\dot{L}_{y}\approx I_{\perp}\ddot{n}_{x}\approx0$. The
high-frequency resonance $\omega_{\perp}$ is blueshifted by $\delta\omega_{\perp}$$\sim$
$\omega_{c}$. As before, the lattice hardly moves in the high-frequency
mode, while at low frequencies the magnetization is locked to the
lattice. 

\begin{figure}
\begin{centering}
\includegraphics{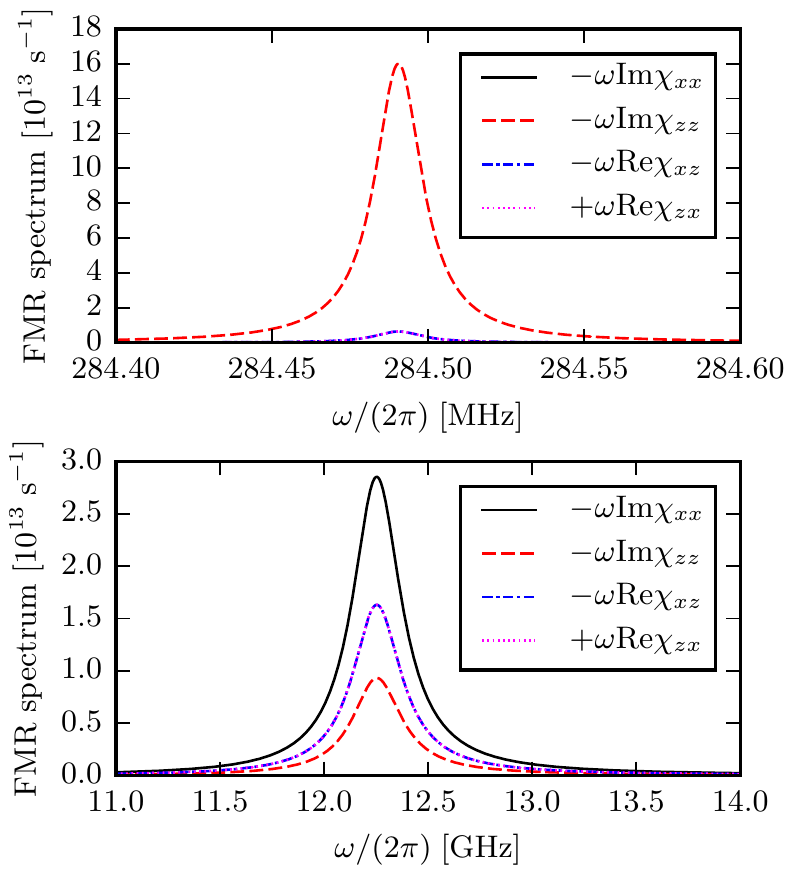}
\par\end{centering}
\caption{FMR spectrum of an Fe disk with $15\;\mathrm{nm}$ diameter and $2\;\mathrm{nm}$
thickness in a static magnetic field of $0.25\;\mathrm{T}$ with Gilbert
damping constant $\alpha=0.01$.\label{fig:FMR disk}}
\end{figure}

In Fig.~\ref{fig:FMR disk} we plot the FMR spectrum of an Fe nanodisk
with shape parameters $a=1\;\mathrm{nm}$ and $b=7.5\;\mathrm{nm}$
under $\mu_{0}H_{0}=0.25\;\mathrm{T}$ or $\omega_{H}/(2\pi)=7\;\mathrm{GHz}$.
The characteristic frequencies are $\omega_{c}/(2\pi)=17.2\;\mathrm{MHz}$
and $\omega_{A}/(2\pi)=-14.4\;\mathrm{GHz}$. The blocking temperature
with $|E_{Z}|/(k_{B}T_{B})\approx24$ is now about $300\;\mathrm{K.}$
Again, the low-frequency resonance is very sharp and relatively weak.
The contribution of $\mathrm{Im}\chi_{xx}$ to the low-frequency resonance
is by a factor of 600 smaller than the dominant $\mathrm{Im}\chi_{zz}$
and therefore not visible in the plot.

\section{Discussion\label{sec:Discussion}}

The examples discussed above safely fulfill all conditions for the
validity of the theory either at reduced temperatures ($T<11\;\mathrm{K}$,
Fe sphere with $2\;\mathrm{nm}$ diameter) or even up to room temperature
($2\;\mathrm{nm}\times15\;\mathrm{nm}$ Fe disk). The levitation of
the particle can be achieved in cluster beams \cite{Bucher91,Billas94,Ma14},
in aerosols \cite{Wang05}, or by confinement to a magnetic trap \cite{Pritchard83,Rusconi16,Prat-Camps17}.
FMR experiments should preferably be carried out in a microwave cavity,
e.g., a coplanar wave guide that can also serve as a trap \cite{Bernon13}.

Metal oxide nanoparticles, such as $\varepsilon$-Fe$_{2}$O$_{3}$
\cite{Ohkoshi15}, have crystal anisotropies of the same order as
that of pure iron but smaller magnetization, which reduces the magnetomechanical
coupling strength, leading to similar results for somewhat smaller
particles. The strongest anisotropies and couplings can be found in
single-molecule magnets, e.g., TbPc$_{2}$ \cite{Ishikawa03}, but
FMR experiments have to be carried out at low temperatures in order
to suppress thermal fluctuations.

Our theory holds for isolated particles at sufficiently low temperatures
and disregards quantum effects. According to the fluctuation-dissipation
theorem a Gilbert damping is at finite temperatures associated with
stochastic fields \cite{Brown63}. A full statistical treatment of
the dynamics of magnetic nanoparticles at elevated temperatures, subject
to microwaves, and weakly coupled to the environment is beyond the
scope of the present paper. When not suspended in vacuum but in, e.g.,
a liquid, the mechanical motion encounters viscous damping and additional
random torques acting on the lattice. Vice versa, the liquid in proximity
of the particle will be stirred by its motion. These effects can be
included in principle by an additional torque term in Eq.~(\ref{eq:mechanical torque}).
The external torque will cause fluctuations in $\Omega_{z}$ and a
temperature dependent broadening of the low-frequency resonances. 

Microwave cavities loaded with thin films or spheres of the high-quality
ferrimagnet yttrium iron garnet have received recent attention because
of the relative ease with which the (ultra) strong coupling between
magnons and photons can be achieved (for references and evidence for
coherent magnon-phonon interaction, see \cite{Zhang16}). The sharp
low-frequency modes of free magnetic nanoparticles coupled to rf cavity
modes at 10-100 MHz correspond to co-operativities that are limited
only by the quality factor of the cavity. This appears to be a promising
route to access non-linear, chaotic, or quantum dynamical regimes.
This technique would work also for magnets with large damping and
could break the monopoly of yttrium iron garnet for quantum cavity
magnonics. Materials with a large anisotropy are most attractive by
the enhanced magnetization-lattice coupling. 

\section{Summary\label{sec:Summary}}

In conclusion, we discussed the effect of the magnetomechanical coupling
on the dynamics of levitated single-domain spheroidal magnetic nanoparticles,
e.g., in molecular cluster beams and aerosols.\textbf{\textcolor{red}{{}
}}We predict a blue shift of the high-frequency resonance and additional
low-frequency satellites in FMR spectra that reflect particle shape
and material parameters. In the low-frequency modes the nanomagnet
precesses together with the magnetization with strongly reduced effective
damping and thereby spectral broadening.
\begin{acknowledgments}
This work is part of the research program of the Stichting voor Fundamenteel
Onderzoek der Materie (FOM), which is financially supported by the
Nederlandse Organisatie voor Wetenschappelijk Onderzoek (NWO) as well
as JSPS KAKENHI Grant Nos. 25247056, 25220910, 26103006. A. K. acknowledges
financial support from the Alexander v. Humboldt foundation. H. K.
would like to express her gratitude toward her late supervisor Malek
Zareyan for the opportunity to collaborate with the TU Delft researchers.
S. S. is grateful to Alejandro O. León for insightful discussions.
\end{acknowledgments}

\appendix

\section{Coordinate systems and transformations\label{sec:Coordinate-systems}}

We derive the coordinate transformation from the lab with basis vectors
$\mathbf{e}_{x}$, $\mathbf{e}_{y}$, $\mathbf{e}_{z}$ to the body
frame $\mathbf{e}_{x_{b}}$, $\mathbf{e}_{y_{b}}$, $\mathbf{e}_{z_{b}}$.
The position of the particle is specified by the three Euler angles
($\phi,\theta,\psi$). These three angles are defined by the transformation
matrix from the lab to the body frame ($\mathbf{r}_{b}=\mathcal{A}\mathbf{r}$),
\begin{eqnarray}
\mathcal{A} & = & \begin{pmatrix}\cos\psi & \sin\psi & 0\\
-\sin\psi & \cos\psi & 0\\
0 & 0 & 1
\end{pmatrix}\begin{pmatrix}1 & 0 & 0\\
0 & \cos\theta & \sin\theta\\
0 & -\sin\theta & \cos\theta
\end{pmatrix}\nonumber \\
 &  & \times\begin{pmatrix}\cos\phi & \sin\phi & 0\\
-\sin\phi & \cos\phi & 0\\
0 & 0 & 1
\end{pmatrix}.
\end{eqnarray}
The main axis $\mathbf{n}$ of the particle is given by the local
$z_{b}$-axis in the body frame and can be directly obtained via the
inverse transformation $\mathcal{A}^{\mathrm{T}}$,
\begin{equation}
\mathbf{n}=\begin{pmatrix}\sin\theta\sin\phi\\
-\sin\theta\cos\phi\\
\cos\theta
\end{pmatrix}.
\end{equation}
The angular velocity vector of the rotating particle reads in the
lab frame
\begin{eqnarray}
\mathbf{{\bf \boldsymbol{\Omega}}} & = & \dot{\psi}\mathcal{A}^{\mathrm{T}}\begin{pmatrix}0\\
0\\
1
\end{pmatrix}+\dot{\theta}\begin{pmatrix}\cos\phi & -\sin\phi & 0\\
\sin\phi & \cos\phi & 0\\
0 & 0 & 1
\end{pmatrix}\begin{pmatrix}1\\
0\\
0
\end{pmatrix}+\dot{\phi}\begin{pmatrix}0\\
0\\
1
\end{pmatrix}\nonumber \\
 & = & \begin{pmatrix}\dot{\theta}\cos\phi+\dot{\psi}\sin\theta\sin\phi\\
\dot{\theta}\sin\phi-\dot{\psi}\sin\theta\cos\phi\\
\dot{\phi}+\dot{\psi}\cos\theta
\end{pmatrix},
\end{eqnarray}
and in the body frame,
\begin{equation}
\mathbf{{\bf \boldsymbol{\Omega}}}_{b}=\mathcal{A}\mathbf{{\bf \boldsymbol{\Omega}}}=\begin{pmatrix}\dot{\phi}\sin\theta\sin\psi+\dot{\theta}\cos\psi\\
\dot{\phi}\sin\theta\cos\psi-\dot{\theta}\sin\psi\\
\dot{\phi}\cos\theta+\dot{\psi}
\end{pmatrix}.
\end{equation}
The mechanical angular momentum $\mathbf{L}$ and the principal axis
$\mathbf{n}$ of the nanomagnet can be related by considering the
mechanical angular momentum in the body frame
\begin{equation}
\mathbf{L}_{b}=\mathcal{I}_{b}\mathbf{{\bf \boldsymbol{\Omega}}}_{b}.\label{eq:angular velocity body frame}
\end{equation}
Transforming (\ref{eq:angular velocity body frame}) to the lab frame
and expanding for small angles $\theta$, \begin{subequations}\label{eq: relation L n}
\begin{eqnarray}
L_{x} & \approx & I_{\perp}\frac{d}{dt}(\theta\cos\phi)\approx-I_{\perp}\dot{n}_{y},\\
L_{y} & \approx & I_{\perp}\frac{d}{dt}(\theta\sin\phi)\approx I_{\perp}\dot{n}_{x},\\
L_{z} & \approx & I_{3}(\dot{\phi}+\dot{\psi})\approx I_{3}\Omega_{z},
\end{eqnarray}
\end{subequations}which is a valid approximation when $\Omega_{z}=\mathcal{O}(\theta).$
Furthermore, $n_{z}\approx1$ and $\dot{n}_{z}\approx0$ is consistent
with $\theta\ll1$.

The Gilbert damping is defined for the relative motion of the magnetization
with respect to the lattice, i.e. in the rotating frame. The damping
in the lab frame is obtained by the coordinate transformation
\begin{equation}
\bm{\tau}_{m}^{(\alpha)}=\mathcal{A}^{\mathrm{T}}\bm{\tau}_{m,b}^{(\alpha)}=\mathcal{A}^{\mathrm{T}}\left(\alpha{\bf m}_{b}\times\dot{{\bf m}}_{b}\right),
\end{equation}
where ${\bf m}_{b}=\mathcal{A}{\bf m}$. Expanding the time derivative
\begin{equation}
\bm{\tau}_{m}^{(\alpha)}=\alpha{\bf m}\times\dot{{\bf m}}+\alpha{\bf m}\times\left(\mathcal{A}^{\mathrm{T}}\dot{\mathcal{A}}{\bf m}\right).
\end{equation}
The angular frequency vector ${\bf \boldsymbol{\Omega}}$ is defined
by
\begin{equation}
\dot{{\bf r}}={\bf {\bf \boldsymbol{\Omega}}}\times{\bf r},\label{eq:rdot1}
\end{equation}
where ${\bf r}$ is a point in the rotating body, i.e. $\dot{{\bf r}}_{b}=0$,
and
\begin{equation}
\dot{{\bf r}}=\dot{\mathcal{A}}^{\mathrm{T}}{\bf r}_{b}=\dot{\mathcal{A}}^{\mathrm{T}}A{\bf r}.\label{eq:rdot2}
\end{equation}
Using $\frac{d}{dt}(\mathcal{A}^{\mathrm{T}}\mathcal{A})=\mathcal{A}^{\mathrm{T}}\dot{\mathcal{A}}+\dot{\mathcal{A}}^{\mathrm{T}}A=0$
and comparing Eqs.~(\ref{eq:rdot1}) and (\ref{eq:rdot2}),
\begin{equation}
\mathcal{A}^{\mathrm{T}}\dot{\mathcal{A}}{\bf r}={\bf r}\times{\bf {\bf \boldsymbol{\Omega}}},
\end{equation}
and therefore
\begin{equation}
\bm{\tau}_{m}^{(\alpha)}=\alpha{\bf m}\times\dot{{\bf m}}+\alpha{\bf m}\times\left({\bf m}\times{\bf \boldsymbol{\Omega}}\right).
\end{equation}

\section{Poisson bracket in Hamilton's equation\label{sec:Hamilton's-equation}}

In the following, we show how to derive Hamilton's equation of motion
(\ref{eq:Hamilton evaluated}). Using the linearity of the Poisson
bracket together with the product rule
\begin{equation}
\left\{ AB,C\right\} =A\left\{ B,C\right\} +\left\{ A,C\right\} B,\label{eq:product rule}
\end{equation}
and $\left\{ a_{ijk}(\mathbf{n},\mathbf{L}),\mathbf{S}\right\} =0$,
we get
\begin{equation}
\left\{ \mathbf{S},\mathcal{H}\right\} =\sum_{i,j,k\in\mathbb{N}_{0}}a_{ijk}(\mathbf{n},\mathbf{L})\left\{ \mathbf{S},S_{x}^{i}S_{y}^{j}S_{z}^{k}\right\} .\label{eq:Hamilton appendix}
\end{equation}
We only consider the $x$-component, as the other components can be
derived similarly. Using the product rule (\ref{eq:product rule}),
we may write

\begin{eqnarray}
\left\{ S_{x},S_{x}^{i}S_{y}^{j}S_{z}^{k}\right\}  & = & S_{x}^{i}\left\{ S_{x},S_{y}^{j}S_{z}^{k}\right\} \nonumber \\
 & = & S_{x}^{i}S_{y}^{j}\left\{ S_{x},S_{z}^{k}\right\} +S_{x}^{i}S_{z}^{k}\left\{ S_{x},S_{y}^{j}\right\} .\nonumber \\
 &  & \,
\end{eqnarray}
Next, we prove by induction that
\begin{equation}
\left\{ S_{x},S_{z}^{k}\right\} =-kS_{y}S_{z}^{k-1},\label{eq:induction}
\end{equation}
where the base case ($k=0$)
\begin{equation}
\left\{ S_{x},S_{z}^{0}\right\} =0
\end{equation}
and the inductive step ($k\to k+1$)
\begin{eqnarray}
\left\{ S_{x},S_{z}^{k+1}\right\}  & = & S_{z}\left\{ S_{x},S_{z}^{k}\right\} +S_{z}^{k}\left\{ S_{x},S_{z}\right\} \nonumber \\
 & = & -(k+1)S_{y}S_{z}^{k}
\end{eqnarray}
complete the proof. Similarly, it follows 
\begin{equation}
\left\{ S_{x},S_{y}^{j}\right\} =jS_{y}^{j-1}S_{z}.
\end{equation}
Summarizing
\begin{eqnarray}
\left\{ S_{x},S_{x}^{i}S_{y}^{j}S_{z}^{k}\right\}  & = & jS_{x}^{i}S_{y}^{j-1}S_{z}^{k+1}\nonumber \\
 & - & kS_{x}^{i}S_{y}^{j+1}S_{z}^{k-1},
\end{eqnarray}
which gives with Eq.~(\ref{eq:Hamilton appendix}) the $x$-component
of Eq.~(\ref{eq:Hamilton evaluated}).

\section{Linearized equations of motion\label{sec:Linearized-equations}}

\subsection{Easy-axis configuration\label{sec:Magnetization-parallel}}

\begin{figure*}
\begin{centering}
\includegraphics{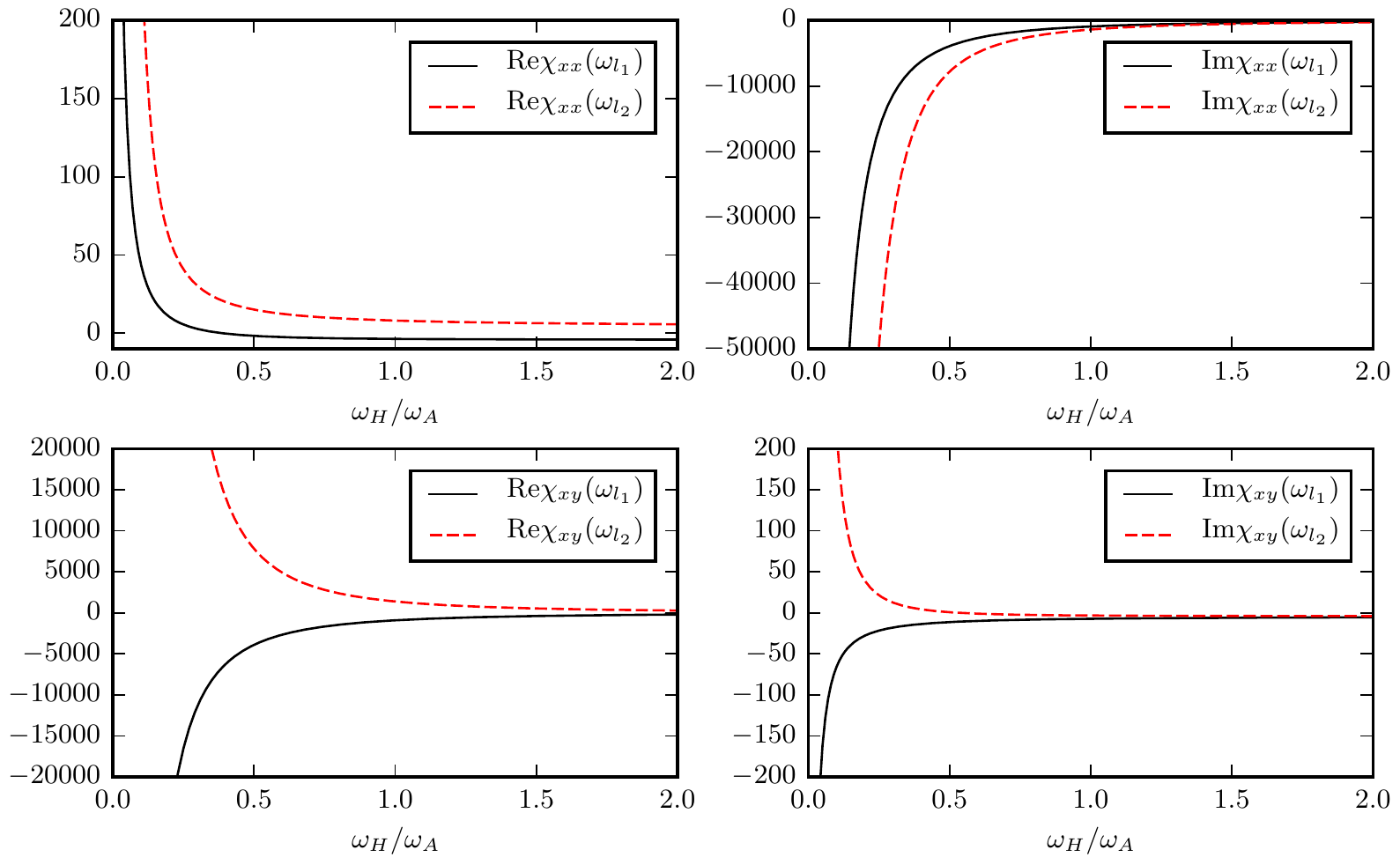}
\par\end{centering}
\caption{Real and imaginary parts of the magnetic susceptibility tensor $\chi(\omega)$
of the low-frequency modes $\omega_{l_{1}}$ and $\omega_{l_{2}}$
for an Fe nanosphere of $2\;\mathrm{nm}$ diameter with Gilbert damping
$\alpha=0.01$.\label{fig:susceptibility}}
\end{figure*}
In the easy-axis case ($\mathbf{m}\parallel\mathbf{n}\parallel\bm{e}_{z}$),
the linearized equations of motion of the magnetization $\mathbf{m}$
and mechanical angular momentum ${\bf L}$ read \begin{subequations}

\begin{eqnarray}
\dot{m}_{x} & = & -\omega_{H}m_{y}+\omega_{M}\frac{h_{y}}{M_{s}}-\omega_{A}\left(m_{y}-n_{y}\right)-\alpha\left(\dot{m}_{y}-\dot{n}_{y}\right),\nonumber \\
\\
\dot{m}_{y} & = & \omega_{H}m_{x}-\omega_{M}\frac{h_{x}}{M_{s}}+\omega_{A}\left(m_{x}-n_{x}\right)+\alpha\left(\dot{m}_{x}-\dot{n}_{x}\right),\nonumber \\
\\
\dot{m}_{z} & = & 0,
\end{eqnarray}
\end{subequations}\begin{subequations}
\begin{eqnarray}
\dot{L}_{x} & = & -I_{\perp}\ddot{n}_{y},\\
\dot{L}_{y} & = & I_{\perp}\ddot{n}_{x},\\
\dot{L}_{z} & = & I_{3}\dot{\Omega}_{z}=0,
\end{eqnarray}
\end{subequations}with\begin{subequations}
\begin{eqnarray}
\ddot{n}_{x} & = & \omega_{N}^{2}\left(m_{x}-n_{x}\right)+\alpha\omega_{c}\left(\dot{m}_{x}-\dot{n}_{x}\right),\\
\ddot{n}_{y} & = & \omega_{N}^{2}\left(m_{y}-n_{y}\right)+\alpha\omega_{c}\left(\dot{m}_{y}-\dot{n}_{y}\right),\\
\ddot{n}_{z} & = & 0,
\end{eqnarray}
\end{subequations}where $\omega_{N}^{2}=\omega_{c}\omega_{A}$. Since
$\dot{\Omega}_{z}=0$ and with initial condition $\Omega_{z}=0$,
there is no net rotation $\Omega_{z}$. Introducing the chiral modes,
\begin{equation}
m^{\pm}=m_{x}\pm im_{y},\;\;\;n^{\pm}=n_{x}\pm in_{y},\;\;\;h^{\pm}=h_{x}\pm ih_{y},
\end{equation}
we can write the equations of motion in the compact form

\begin{eqnarray}
\dot{m}^{\pm} & = & \pm i\left(\omega_{0}m^{\pm}-\omega_{M}\frac{h^{\pm}}{M_{s}}-\omega_{A}n^{\pm}\right)\pm i\alpha\left(\dot{m}^{\pm}-\dot{n}^{\pm}\right),\nonumber \\
\\
\ddot{n}^{\pm} & = & \omega_{N}^{2}\left(m^{\pm}-n^{\pm}\right)+\alpha\omega_{c}\left(\dot{m}^{\pm}-\dot{n}^{\pm}\right).
\end{eqnarray}
For ac magnetic fields
\begin{equation}
h^{\pm}(t)=h_{0}^{\pm}e^{i\omega t},
\end{equation}
we solve the equations of motion by the ansatz
\begin{equation}
m^{\pm}(t)=m_{0}^{\pm}e^{i\omega t},\;\;\;n^{\pm}(t)=n_{0}^{\pm}e^{i\omega t}.
\end{equation}
The observables correspond to the real part of the complex $\mathbf{m}$,
$\mathbf{n},$ and $\mathbf{h}$. The susceptibilities are defined
\begin{equation}
m^{\pm}=\chi^{\pm}h^{\pm}/M_{s},\;\;\;n^{\pm}=\chi_{n}^{\pm}m^{\pm},
\end{equation}
and read 

\begin{equation}
\chi_{n}^{\pm}(\omega)=\frac{\omega_{N}^{2}+i\alpha\omega\omega_{c}}{-\omega^{2}+\omega_{N}^{2}+i\alpha\omega\omega_{c}},\label{eq:susceptibility nm}
\end{equation}
\begin{eqnarray}
\chi^{\pm}(\omega) & = & \mp\omega_{M}(-\omega^{2}+\omega_{N}^{2}+i\alpha\omega\omega_{c})\nonumber \\
 & \times & \left[(\omega\mp\omega_{0}\mp i\alpha\omega)(-\omega^{2}+\omega_{N}^{2}+i\alpha\omega\omega_{c})\right.\nonumber \\
 &  & \left.\pm\omega_{c}(\omega_{A}+i\alpha\omega)^{2}\right]^{-1}.\label{eq:susceptibility parallel}
\end{eqnarray}
Close to a resonance of $\chi^{\pm}$ at $\omega_{i}$ the absorbed
microwave power is determined by the contributions
\begin{equation}
-\frac{\omega}{2}\mathrm{Im}\chi^{\pm}(\omega)\sim F^{\pm}\frac{\left(\eta^{\pm}\right)^{2}}{(\omega-\omega_{i})^{2}+\left(\eta^{\pm}\right)^{2}},
\end{equation}
with 
\begin{equation}
\eta^{\pm}=\frac{\pm\alpha\omega_{i}\left(\omega_{i}^{2}+\omega_{c}(\pm\omega_{i}-\omega_{H})\right)}{3\omega_{i}^{2}\mp2\omega_{i}\omega_{0}-\omega_{c}\omega_{A}},
\end{equation}

\begin{equation}
F^{\pm}=\frac{\frac{1}{2}\omega_{M}(\omega_{i}^{2}-\omega_{c}\omega_{A})}{\alpha\left(\omega_{i}^{2}+\omega_{c}(\pm\omega_{i}-\omega_{H})\right)}.
\end{equation}
Note that for each resonance of $\chi^{+}$ at $\omega_{i}$ there
is a corresponding resonance of $\chi^{-}$ at $-\omega_{i}$. 

The magnitudes of the $x$- and $y$-components of $\mathbf{n}$ are
related to $\mathbf{m}$ via the susceptibility $\chi_{n}^{\pm}$
given in Eq.~(\ref{eq:susceptibility nm}). For high frequencies
$\omega$ we find $\chi_{n}^{\pm}\approx0$ and for low frequencies
$\chi_{n}^{\pm}\approx1$. Therefore, the main axis $\mathbf{n}$
is nearly static for the high-frequency mode, while for the low-frequency
modes $\mathbf{n}$ stays approximately parallel to $\mathbf{m}$.

The susceptibility $\chi^{\pm}$ given in Eq.~(\ref{eq:susceptibility parallel})
can be related to the usual magnetic susceptibilities $(\alpha,\beta=x,y$),
\begin{equation}
\chi_{\alpha\beta}=\left.\frac{M_{\alpha}}{h_{\beta}}\right|_{\mathbf{h}_{\perp}=0}.
\end{equation}
Defining the symmetric and antisymmetric parts of the susceptibility
$\chi^{\pm}$,
\begin{equation}
\chi^{\pm}=\chi_{s}\pm\chi_{a}.\label{eq:chi s/a}
\end{equation}
we find the relations \begin{subequations}

\begin{eqnarray}
\chi_{xx} & = & \chi_{yy}=\chi_{s},\\
\chi_{xy} & = & -\chi_{yx}=i\chi_{a}.
\end{eqnarray}
\end{subequations}

The magnetization dynamics in terms of the magnetic susceptibility
reads
\begin{equation}
\mathrm{Re}\begin{pmatrix}m_{x}(t)\\
m_{y}(t)
\end{pmatrix}=\mathrm{Re}\left[\begin{pmatrix}\chi_{xx} & \chi_{xy}\\
-\chi_{xy} & \chi_{xx}
\end{pmatrix}\begin{pmatrix}h_{x}(t)/M_{s}\\
h_{y}(t)/M_{s}
\end{pmatrix}\right],
\end{equation}
where $\chi_{yy}=\chi_{xx}$ and $\chi_{yx}=-\chi_{xy}$. For linear
polarization $h_{x}(t)=|h_{x}|e^{i\omega t}$ and $h_{y}(t)=0$, 
\begin{equation}
\mathrm{Re}\begin{pmatrix}m_{x}(t)\\
m_{y}(t)
\end{pmatrix}=\frac{|h_{x}|}{M_{s}}\begin{pmatrix}\mathrm{Re}\chi_{xx}\cos(\omega t)-\mathrm{Im}\chi_{xx}\sin(\omega t)\\
-\mathrm{Re}\chi_{xy}\cos(\omega t)+\mathrm{Im}\chi_{xy}\sin(\omega t)
\end{pmatrix}.
\end{equation}
According to Fig.~\ref{fig:susceptibility}, $|\mathrm{Re}\chi_{xx}|,\,|\mathrm{Im}\chi_{xy}|\ll|\mathrm{Re}\chi_{xy}|\approx|\mathrm{Im}\chi_{xx}|$,
and $\mathrm{Im}\chi_{xx}<0$ for both low-frequency modes $\omega_{l_{1}}$
and $\omega_{l_{2}}$. The direction of the precession depends now
on the sign of $\mathrm{Re}\chi_{xy}$, which is negative for $\omega_{l_{1}}$
and positive for $\omega_{l_{2}}$. The mode $\omega_{l_{1}}$ is
a clockwise precession,
\begin{equation}
\mathrm{Re}\begin{pmatrix}m_{x}(t)\\
m_{y}(t)
\end{pmatrix}\propto\begin{pmatrix}\sin(\omega_{l_{1}}t)\\
\cos(\omega_{l_{1}}t)
\end{pmatrix},
\end{equation}
whereas the mode $\omega_{l_{2}}$ precesses counterclockwise:
\begin{equation}
\mathrm{Re}\begin{pmatrix}m_{x}(t)\\
m_{y}(t)
\end{pmatrix}\propto\begin{pmatrix}\sin(\omega_{l_{2}}t)\\
-\cos(\omega_{l_{2}}t)
\end{pmatrix}.
\end{equation}
Note that $\chi^{-}(\omega)$ has a low-frequency peak only at $\omega_{l_{1}}$
and $\chi^{+}(\omega)$ only at $\omega_{l_{2}}$ (for $\omega>0$). 

\subsection{Easy-plane configuration\label{sec:Magnetization-normal}}

Here, we consider an equilibrium magnetization normal to the principal
axis ($\mathbf{m}\perp\mathbf{n}$) due to the shape anisotropy of
an oblate spheroid. Linearizing for small deviations from the equilibrium
($\theta\ll1$, $m_{y}\approx1$, $n_{z}\approx1$), the equations
of motion for the magnetization and mechanical angular momentum read\begin{subequations}

\begin{eqnarray}
\dot{m}_{x} & = & \omega_{H}m_{z}-\omega_{M}\frac{h_{z}}{M_{s}}-\omega_{A}\left(m_{z}+n_{y}\right)+\alpha\left(\dot{m}_{z}+\dot{n}_{y}\right),\nonumber \\
\label{eq:mx_dot normal}\\
\dot{m}_{y} & = & 0,\\
\dot{m}_{z} & = & -\omega_{H}m_{x}+\omega_{M}\frac{h_{x}}{M_{s}}-\alpha\dot{m}_{x}-\alpha\Omega_{z},\label{eq:mz_dot normal}
\end{eqnarray}
\end{subequations}\begin{subequations}
\begin{eqnarray}
\dot{L}_{x} & = & -I_{\perp}\ddot{n}_{y},\\
\dot{L}_{y} & = & I_{\perp}\ddot{n}_{x},\\
\dot{L}_{z} & = & I_{3}\dot{\Omega}_{z}=\frac{VM_{s}}{\gamma}\left(-\alpha\dot{m}_{x}-\alpha\Omega_{z}\right),\label{eq:Lz_dot dormal}
\end{eqnarray}
\end{subequations}with\begin{subequations}
\begin{eqnarray}
\ddot{n}_{x} & = & 0,\\
\ddot{n}_{y} & = & \omega_{N}^{2}\left(m_{z}+n_{y}\right)-\alpha\omega_{c}\left(\dot{m}_{z}+\dot{n}_{y}\right),\label{eq:ny mz}\\
\ddot{n}_{z} & = & 0.
\end{eqnarray}
\end{subequations}In the presence of ac magnetic fields
\begin{equation}
h_{x}(t)=h_{x,0}e^{i\omega t},\;\;\;h_{z}(t)=h_{z,0}e^{i\omega t},
\end{equation}
 we use the ansatz
\begin{equation}
m_{x}(t)=m_{x,0}e^{i\omega t},\;\;\;m_{z}(t)=m_{z,0}e^{i\omega t},\;\;\;n_{y}(t)=n_{y,0}e^{i\omega t}.
\end{equation}
From Eq.~(\ref{eq:Lz_dot dormal}) 
\begin{equation}
\Omega_{z}=\frac{-\omega_{I}\omega\alpha m_{x}}{\omega-i\alpha\omega_{I}}\approx-\alpha\omega_{I}m_{x},\label{eq:Omega_z approximation}
\end{equation}
where $\omega_{I}=VM_{s}/(\gamma I_{3})$ and provided $\alpha\omega_{I}$
is sufficiently smaller than all the other relevant frequencies. We
approximate $\alpha\Omega_{z}=\mathcal{O}(\alpha^{2})\approx0$ in
Eq.~(\ref{eq:mz_dot normal}). Due to the reduced symmetry for $\mathbf{m}\perp\mathbf{n}$,
we cannot simplify the equations of motion by introducing chiral modes,
but have to calculate the Cartesian components of the magnetic susceptibility
tensor $\chi$ as \begin{subequations}\label{eq:chi tensor}
\begin{eqnarray}
\chi_{xx} & = & \omega_{M}\left[\omega^{2}(\omega_{A}-\omega_{H})-i\alpha(\omega^{3}-\omega\omega_{c}\omega_{H})-\omega_{H}\omega_{N}^{2}\right]/\chi_{d},\nonumber \\
\\
\chi_{zz} & = & -\omega_{M}(\omega_{H}+i\alpha\omega)(\omega^{2}+\omega_{N}^{2}-i\alpha\omega_{c}\omega)/\chi_{d},\\
\chi_{xz} & = & i\omega\omega_{M}(\omega^{2}+\omega_{N}^{2}-i\alpha\omega_{c}\omega)/\chi_{d},\\
\chi_{zx} & = & -\chi_{xz},
\end{eqnarray}
\end{subequations}where the denominator 
\begin{eqnarray}
\chi_{d} & = & \omega^{4}(1+\alpha^{2})+i\alpha\omega^{3}(\omega_{A}-\omega_{c}-2\omega_{H})\nonumber \\
 & + & \omega^{2}(\omega_{A}\omega_{H}-\omega_{H}^{2}+\omega_{N}^{2}-\alpha^{2}\omega_{c}\omega_{H})\nonumber \\
 & + & i\alpha\omega\omega_{H}(\omega_{c}\omega_{H}-\omega_{N}^{2})-\omega_{H}^{2}\omega_{N}^{2}.
\end{eqnarray}
The singularities in $\chi$ mark the two resonance frequencies. For
small damping ($\alpha\ll1$)
\begin{eqnarray}
\omega_{1,2}^{2} & = & -\frac{1}{2}(\omega_{A}\omega_{H}-\omega_{H}^{2}+\omega_{N}^{2})\nonumber \\
 &  & \pm\frac{1}{2}\sqrt{(\omega_{A}\omega_{H}-\omega_{H}^{2}+\omega_{N}^{2})^{2}+4\omega_{H}^{2}\omega_{N}^{2}}.
\end{eqnarray}
From Eq.~(\ref{eq:ny mz}), we obtain the following relation between
the magnetic and mechanical motion
\begin{equation}
n_{y}=\frac{-\omega_{N}^{2}+i\alpha\omega_{c}\omega}{\omega^{2}+\omega_{N}^{2}-i\alpha\omega_{c}\omega}m_{z}.
\end{equation}
For high frequencies $n_{y}\approx0$ and for low frequencies $n_{y}\approx-m_{z}$.
This implies that for the high frequency mode $\omega_{\perp}=\omega_{1}$
we recover the bulk FMR, while in the low-frequency mode $\omega_{l}=\omega_{2}$
the magnetization is locked to the lattice. 

\section{FMR absorption\label{sec:FMR-absorption}}

FMR absorption spectra are proportional to the energy dissipated in
the magnet \cite{Labanowski16}. The energy density of the magnetic
field is given by
\begin{equation}
w(t)=\frac{1}{2}\mathbf{H}(t)\cdot\mathbf{B}(t),
\end{equation}
where $\mathbf{B}=\mu_{0}\chi\mathbf{H}$. The absorbed microwave
power by a magnet of volume $V$ is
\begin{equation}
P(t)=V\dot{w}(t)=V\mathbf{H}(t)\cdot\dot{\mathbf{B}}(t).
\end{equation}
The average over one cycle $T=2\pi/\omega$,
\begin{equation}
P\equiv\left\langle P(t)\right\rangle =\frac{1}{T}\int_{0}^{T}dt\,P(t),
\end{equation}
can be calculated using the identity
\begin{equation}
\left\langle \mathrm{Re}(\mathbf{A}e^{i\omega t})\cdot\mathrm{Re}(\mathbf{B}e^{i\omega t})\right\rangle =\frac{1}{2}\mathrm{Re}\left(\mathbf{A}^{*}\cdot\mathbf{B}\right).
\end{equation}
When a monochromatic ac component of the magnetic field $\mathbf{h}_{\perp}$
is normal to its dc component, the power reads 
\begin{equation}
P=-\frac{\mu_{0}V}{2}\omega\mathrm{Im}\left(\mathbf{h}_{\perp}^{*}\cdot\mathbf{M}_{\perp}\right),
\end{equation}
where $\mathbf{M}_{\perp}$ is the transverse magnetization. When
the magnetization and static magnetic field are parallel to the principal
axis of the particle, we can write 

\begin{eqnarray}
P & = & -\frac{\mu_{0}V}{2}\omega\left[(\left|h_{x}\right|^{2}+\left|h_{y}\right|^{2})\mathrm{Im}\chi_{s}(\omega)\right.\nonumber \\
 &  & -\left.2\mathrm{Im}(h_{x}^{*}h_{y})\mathrm{Im}\chi_{a}(\omega)\right],
\end{eqnarray}
where the symmetric and antisymmetric parts of the susceptibility
$\chi^{\pm}$ Eq.~(\ref{eq:susceptibility parallel}) as defined
by Eq.~(\ref{eq:chi s/a}) obey the symmetry relations $\mathrm{Im}\chi_{s}(-\omega)=-\mathrm{Im}\chi_{s}(\omega)$
and $\mathrm{Im}\chi_{a}(-\omega)=\mathrm{Im}\chi_{a}(\omega)$. The
term proportional to $\mathrm{Im}\chi_{a}$ can therefore be negative,
depending on the signs of $\omega$ and $\mathrm{Im}(h_{x}^{*}h_{y})$,
whereas the term involving $\mathrm{Im}\chi_{s}$ (as well as the
total absorbed power) is always positive. 

When magnetization and static magnetic field are normal to the principal
axis, both real and imaginary parts of the off-diagonal components
of $\chi$ contribute to the absorbed power
\begin{eqnarray}
P & = & -\frac{\mu_{0}V}{2}\omega\left[|h_{x}|^{2}\mathrm{Im}\chi_{xx}(\omega)+|h_{z}|^{2}\mathrm{Im}\chi_{zz}(\omega)\right.\nonumber \\
 &  & +\left.\mathrm{Im}(\chi_{xz}h_{x}^{*}h_{z}+\chi_{zx}h_{x}h_{z}^{*})\right].
\end{eqnarray}

\bibliographystyle{apsrev4-1}

\end{document}